\documentclass[review]{elsarticle}

\usepackage{lineno,hyperref}
\usepackage{graphicx}
\usepackage{amsmath}
\usepackage{siunitx}
\usepackage{float}
\usepackage{gensymb}
\modulolinenumbers[5]

\begin{document}

\begin{frontmatter}

\title{Angle-dependent electron spin resonance of $\mathrm{YbRh}_2\mathrm{Si}_2$ measured with planar microwave resonators and in-situ rotation}

\author[address1]{Linda Bondorf}
\author[address1]{Manfred Beutel}
\author[address1]{Markus Thiemann}
\author[address1]{Martin Dressel}
\author[address1a]{Daniel Bothner\fnref{fn1}}
\author[address3]{J{\"o}rg Sichelschmidt}
\author[address2]{Kristin Kliemt}
\author[address2]{Cornelius Krellner}
\author[address1]{Marc Scheffler\corref{mycorrespondingauthor}}
\ead[url]{scheffl@pi1.physik.uni-stuttgart.de}
\cortext[mycorrespondingauthor]{Corresponding author}
\fntext[fn1]{Present address: Kavli Institute of NanoScience, Delft University of Technology, Delft, The Netherlands}
\address[address1]{1. Physikalisches Institut, Universit{\"a}t Stuttgart, Germany}
\address[address1a]{Physikalisches Institut and Center for Quantum Science (CQ) in LISA+, Universit{\"a}t T{\"u}bingen, Germany}
\address[address3]{Max-Planck-Institut f{\"u}r Chemische Physik fester Stoffe, Dresden, Germany}\address[address2]{Goethe-Universit{\"a}t, Frankfurt am Main, Germany}

\begin{abstract}
We present a new experimental approach to investigate the magnetic properties of the anisotropic heavy-fermion system $\mathrm{YbRh}_2\mathrm{Si}_2$ as a function of crystallographic orientation. Angle-dependent electron spin resonance (ESR) measurements are performed at a low temperature of \SI{1.6}{K} and at an ESR frequency of \SI{4.4}{GHz} utilizing a superconducting planar microwave resonator in a $^4$He-cryostat in combination with in-situ sample rotation. 
The obtained ESR \textit{g}-factor of $\mathrm{YbRh}_2\mathrm{Si}_2$ as a function of the crystallographic angle is consistent with results of previous measurements using conventional ESR spectrometers at higher frequencies and fields. Perspectives to implement this experimental approach into a dilution refrigerator and to reach the magnetically ordered phase of $\mathrm{YbRh}_2\mathrm{Si}_2$ are discussed.
\end{abstract}


\end{frontmatter}

\newpage
\section{Introduction}
The tetragonal heavy-fermion metal $\mathrm{YbRh}_2\mathrm{Si}_2$ shows pronounced magnetic anisotropy \cite{Trovarelli2000a, Trovarelli2000b, gegenwart2002} and is an intensively studied model system for quantum criticality \cite{gegenwart2008}. It exhibits antiferromagnetic order at temperatures below \SI{70}{mK} and in-plane magnetic fields below \SI{60}{mT} \cite{gegenwart2002}. Its N\'eel temperature $T_N$ decreases with increasing field down to a quantum-critical point (with $T_N=0$) induced by the external magnetic field of \SI{60}{mT} (within the tetragonal $ab$-plane) or \SI{660}{mT} (along the $c$-axis) \cite{gegenwart2008,custers2003}. Due to the presence of the quantum-critical point, the system shows pronounced non-Fermi-liquid properties \cite{gegenwart2008,custers2003}.
As the antiferromagnetic state underlies the quantum-critical nature of $\mathrm{YbRh}_2\mathrm{Si}_2$, the details of the magnetic order are highly interesting in context of the peculiar properties of $\mathrm{YbRh}_2\mathrm{Si}_2$. However, due to major experimental challenges in commonly used methods such as neutron scattering \cite{stock2012}, the magnetically ordered phase of $\mathrm{YbRh}_2\mathrm{Si}_2$ is not sufficiently investigated and understood yet. ESR could be a promising alternative method to elucidate details of the antiferromagnetic order, but conventional ESR spectrometers are limited in both temperature and magnetic field to energies much higher than the magnetic order of $\mathrm{YbRh}_2\mathrm{Si}_2$. 
Multiple ESR investigations on $\mathrm{YbRh}_2\mathrm{Si}_2$ have been performed \cite{sichelschmidt2003,sichelschmidt2007, wykhoff2007,Duque2009,schaufuss2009}, but they could not reach the mK temperature range that is required to address the regimes that are key to understanding the quantum-critical nature of $\mathrm{YbRh}_2\mathrm{Si}_2$. As the ESR response of $\mathrm{YbRh}_2\mathrm{Si}_2$ is a very interesting topic on its own \cite{sichelschmidt2003} and as its possible relation to quantum criticality is not settled \cite{Kochelaev2009,Woelfle2009}, ESR studies close to the quantum-critical point are also desired from a fundamental perspective of magnetic resonance.
Planar microwave resonators can be used as ESR probes \cite{scheffler2013,Javaheri2016,Ghirri2015} for $\mathrm{YbRh}_2\mathrm{Si}_2$ to overcome the limitations of conventional ESR spectrometers: as such resonators \cite{Frunzio2005,Goeppl2008,Clauss2013,Malissa2013} can be operated with a multimode measurement technique \cite{scheffler2013,DiIorio1988,Hafner2014,Thiemann2017}, they can simultaneously address multiple ESR frequencies and thus multiple magnetic magnetic fields in the phase diagram \cite{scheffler2013}, and they can also be employed at mK temperatures \cite{Thiemann2017,Frunzio2005,wiemann2015,Scheffler2015,Parkkinen2015,Voesch2015}. 

\begin{figure}[tbh]
\centering
\includegraphics[width=0.99\textwidth]{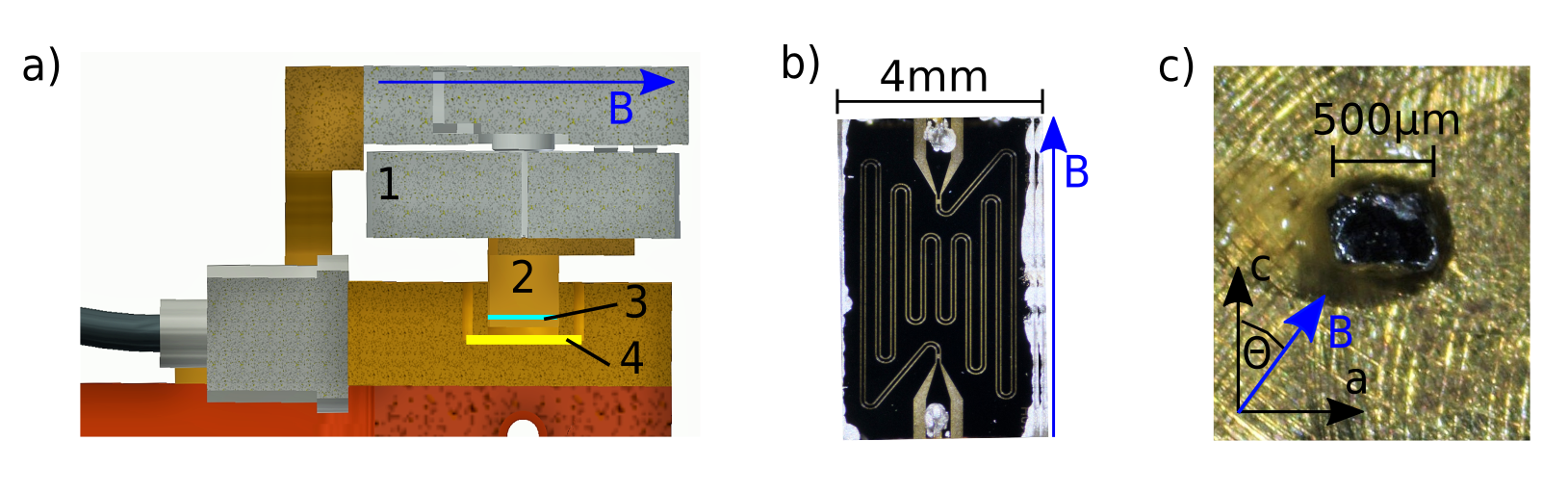}
\caption{Probe and sample mounting for ESR measurements and orientation of static magnetic field $B$. a) Construction inside the sample box with piezoelectric rotator (1), brass stamp (2), sample (3) and coplanar microwave resonator (4).  b) Superconducting Nb resonator on sapphire substrate, with fundamental frequency of \SI{1.5}{GHz}. c) Sample of $\mathrm{YbRh}_2\mathrm{Si}_2$ on a brass stamp, with angle $\Theta$ indicating the orientation of magnetic field $B$ with respect to crystallographic axes $a$ and $c$.}
\label{pic:1}
\end{figure}

\section{Experiment}
We performed microwave measurements on $\mathrm{YbRh}_2\mathrm{Si}_2$ inside a $^4$He-cryostat equipped with a superconducting electromagnet. The arrangement of microwave probe and sample is shown in Fig.\ \ref{pic:1}a): the flat $\mathrm{YbRh}_2\mathrm{Si}_2$ sample is kept at a small distance parallel to the microwave resonator chip (see Fig.\ \ref{pic:1}a)) and is mounted (see Fig.\ \ref{pic:1}c)) via a brass stamp to a commercial piezoelectric rotator  \cite{Attocube}, and thus can be rotated within the sample plane. The microwave chip with meander-type superconducting Nb resonator (see Fig.\ \ref{pic:1}b)) is installed in a brass box and connected via coaxial cables to the 
vector network analyzer for microwave transmission measurements.

In this arrangement of resonator and sample, precise ESR measurements on high-quality $\mathrm{YbRh}_2\mathrm{Si}_2$ single cystals \cite{krellner2012} require sample dimensions of order one millimeter for the relevant surface, namely the $ac$-plane (or another plane that includes the $c$-axis). While such sample dimensions can readily be obtained for the $ab$-plane, growing a $\mathrm{YbRh}_2\mathrm{Si}_2$ crystal with millimeter dimension along the $c$-direction is challenging. This work is the first demonstration of an ESR measurement that combines a planar resonator and a $\mathrm{YbRh}_2\mathrm{Si}_2$ sample in $ac$-plane.

\begin{figure}[tbh]
\centering
\includegraphics[width=0.95\textwidth]{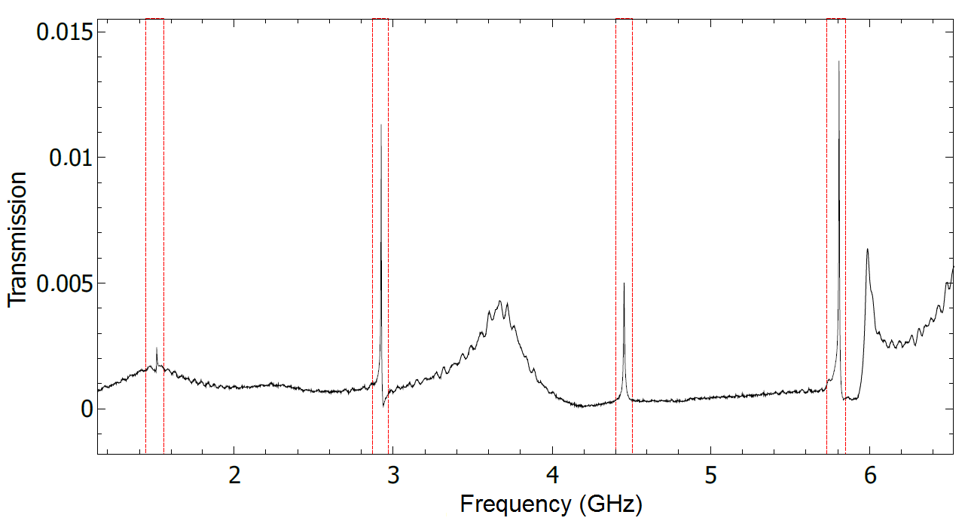}
\caption{Transmission spectrum of the superconducting Nb resonator with equidistant resonance peaks. Frequency ranges that are used to investigate the resonances are indicated. Additional features in the transmission might be due to box modes of the resonator mounting.}
\label{pic:2}
\end{figure}

\begin{figure}[tbh]
\centering
\includegraphics[width=0.85\textwidth]{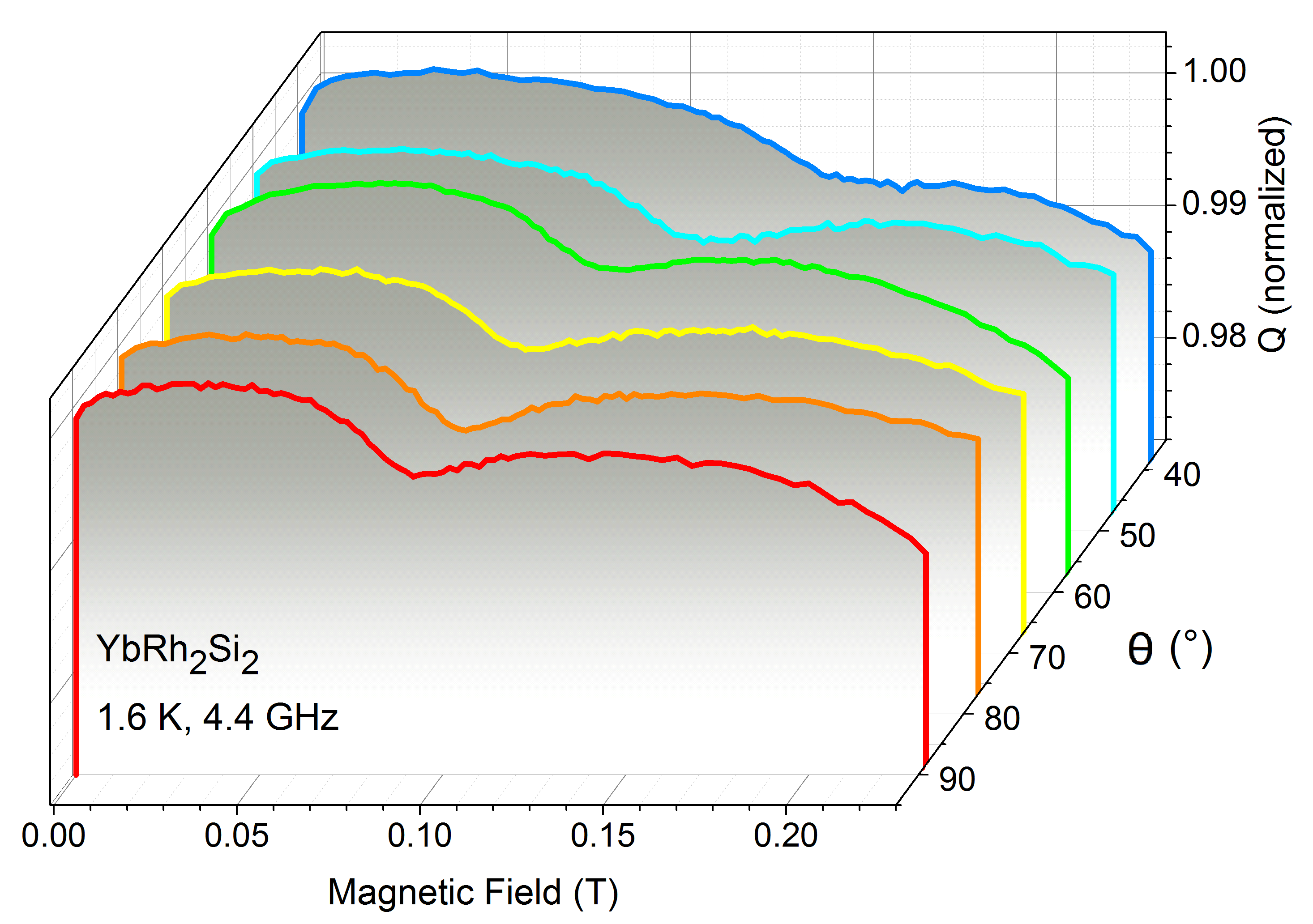}
\caption{ESR in $\mathrm{YbRh}_2\mathrm{Si}_2$ at \SI{1.6}{K} and \SI{4.4}{GHz}. The quality factor $Q$ as a function of external magnetic field and normalized to the zero-field value, shown for several angles $\Theta$ between the external field and the symmetry axis $c$ of the crystal, exhibits a pronounced minimum at the ESR field $B_0$.}
\label{pic:5}
\end{figure}

\section{Results}
Fig.\ \ref{pic:2} shows a typical transmission spectrum of the microwave resonator, and it clearly features sharp resonances of the first four harmonics with roughly equidistant resonance frequencies around \SI{1.5}, \SI{2.9}, \SI{4.4}, and \SI{5.8}{GHz}.
From such spectra, we determine the quality factor $Q_n$ of the $n$-th mode through a Lorentzian fit of the resonance peak in the transmission signal. $Q_n$ is defined as the ratio between resonance frequency $f_n$ and bandwidth (resonance width at half maximum) $\Delta f_n$:
\begin{equation}
Q_n = \dfrac{f_n}{\Delta f_n}
\end{equation}
The transmission in the frequency range close to a resonance peak is measured as a function of magnetic field, and a pronounced change of the resonance peak can be observed. Thus we obtain the field dependence of the quality factor, which is shown in Fig.\ \ref{pic:5} for different sample orientations. $Q$ is generally decreasing with increasing field due to field-induced losses in the superconducting resonator \cite{Frunzio2005,Bothner2012a,deGraaf2012,Bothner2012b,Ebensperger2016} and due to the microwave charge response of the heavy fermions \cite{scheffler2013,Parkkinen2015,Degiorgi1999,Scheffler2005c,Scheffler2010}, and additionally it has a dip at the resonance magnetic field $B_{0}$, indicating ESR absorption and thus the spin response of $\mathrm{YbRh}_2\mathrm{Si}_2$. As can be seen in Fig.\ \ref{pic:5}, the resonance field $B_{0}$ increases if the angle $\Theta$ between the external magnetic field and the crystallographic $c$-axis is decreased from 90$\degree$ ($ab$-plane) towards the $c$-axis.

\begin{figure}[tbh]
\centering
\includegraphics[width=0.9\textwidth]{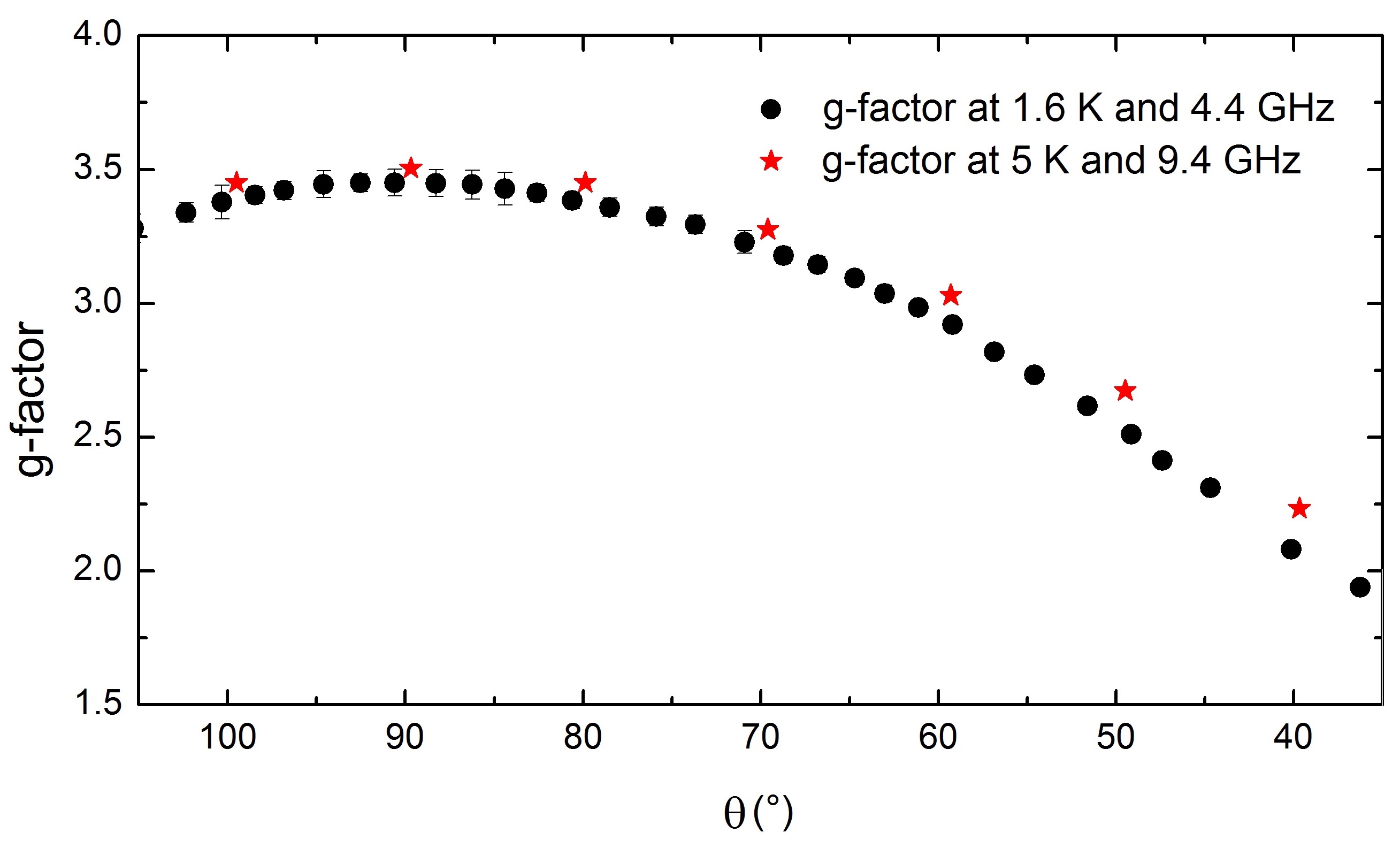}
\caption{ESR $g$-factor  for $\mathrm{YbRh}_2\mathrm{Si}_2$ as a function of the crystallographic angle $\Theta$ measured by coplanar resonator at \SI{1.6}{K} and \SI{4.4}{GHz} (circles). Previous results obtained with a conventional X-band ESR spectrometer at \SI{5}{K} and \SI{9.4}{GHz} are shown for comparison (stars) \cite{sichelschmidt2007}.}
\label{pic:3}
\end{figure}

After subtraction of the background the resonance magnetic field $B_0$ and ESR linewidth $\Delta B$ can be obtained through a fit of an inverse Dysonian function, which describes the absorbed power $P(B)$ at ESR in dependence of the magnetic field strength \cite{joshi2004} 
\begin{equation}
P(B)= \dfrac{\Delta B + \alpha (B-B_0) }{4(B-B_0)^2+\Delta B^2}+ \dfrac{\Delta B - \alpha (B+B_0) }{4(B+B_0)^2+\Delta B^2}.
\end{equation}
The obtained $B_0$ is inserted into the ESR resonance condition to determine the ESR $g$-factor (with Planck constant $h$ and Bohr magneton $\mu_{\mathrm{B}}$):
\begin{equation}
g = \dfrac{h \nu}{B_0 \mu_{\mathrm{B}}}
\end{equation}
Fig.\ \ref{pic:3} shows the ESR $g$-factor of the angle-dependent measurement at \SI{1.6}{K} and \SI{4.4}{GHz} between $\Theta=104^{\circ}$ and $\Theta=36^{\circ}$. There is a maximum around $\Theta = 90^{\circ}$, and $g$ continuously decreases when the crystal is rotated such that the orientation of the magnetic field moves from the $ab$-plane towards the $c$-axis. This evolution is consistent with data obtained previously at higher temperatures (\SI{5}{K}) and higher field with a conventional X-band spectrometer \cite{sichelschmidt2007}, which are shown as stars in Fig.\ \ref{pic:3} for comparison. The slightly higher absolute values of $g$ for the X-band measurement can be explained by the well-established decrease of the $g$-factor with decreasing temperature \cite{sichelschmidt2003,scheffler2013}.

\section{Conclusions}
The presented results prove that planar resonators are sensitive enough for ESR measurements of $\mathrm{YbRh}_2\mathrm{Si}_2$ for magnetic-field directions within the $ac$-plane. We performed the first ESR measurement using a superconducting coplanar microwave resonator in combination with in-situ sample rotation, and this also includes the lowest temperature of any angle-dependent ESR measurement on $\mathrm{YbRh}_2\mathrm{Si}_2$ so far. Our results agree well with former results obtained by conventional techniques at higher frequencies and higher temperatures.
\\
To investigate the magnetically ordered phase of $\mathrm{YbRh}_2\mathrm{Si}_2$ below \SI{70}{mK} and \SI{60}{mT}, this experimental approach for angle-dependent ESR measurements can now be implemented inside a dilution refrigerator. There the technical situation features additional challenges due to the smaller available cooling power (to be considered for the wiring and electrical power needed to drive the rotator). On the other hand, the generally increasing ESR intensity upon cooling promises strong signal for such an experiment.

\section*{Acknowledgments}
We thank G. Untereiner for support during preparation of experiments and R. Kleiner and D. Koelle for support during resonator fabrication. We thank the Deutsche Forschungsgemeinschaft (DFG, projects KR 3831/4-1, SCHE 1580/2-1, SI 1339/1-1) for financial support of this project.

\section*{References}

\bibliography{literature_2017-06-28}

\begin{thebibliography}{10}
\expandafter\ifx\csname url\endcsname\relax
  \def\url#1{\texttt{#1}}\fi
\expandafter\ifx\csname urlprefix\endcsname\relax\def\urlprefix{URL }\fi
\expandafter\ifx\csname href\endcsname\relax
  \def\href#1#2{#2} \def\path#1{#1}\fi

\bibitem{Trovarelli2000a}
O.~Trovarelli, C.~Geibel, S.~Mederle, C.~Langhammer, F.~M. Grosche,
  P.~Gegenwart, M.~Lang, G.~Sparn, F.~Steglich, {YbRh}$_2${Si}$_2$: Pronounced
  non-fermi-liquid effects above a low-lying magnetic phase transition, Phys.
  Rev. Lett. 85 (2000) 626--629.
\newblock \href {http://dx.doi.org/10.1103/PhysRevLett.85.626}
  {\path{doi:10.1103/PhysRevLett.85.626}}.

\bibitem{Trovarelli2000b}
O.~Trovarelli, C.~Geibel, C.~Langhammer, S.~Mederle, P.~Gegenwart, F.~Grosche,
  M.~Lang, G.~Sparn, F.~Steglich, Non-fermi-liquid effects at ambient pressure
  in the stoichiometric heavy-fermion compound {YbRh}$_2${Si}$_2$, Physica B:
  Condensed Matter 281 (2000) 372--373.
\newblock \href {http://dx.doi.org/10.1016/S0921-4526(99)01124-2}
  {\path{doi:10.1016/S0921-4526(99)01124-2}}.

\bibitem{gegenwart2002}
P.~Gegenwart, J.~Custers, C.~Geibel, K.~Neumaier, T.~Tayama, K.~Tenya,
  O.~Trovarelli, F.~Steglich, Magnetic-field induced quantum critical point in
  {YbRh}$_2${Si}$_2$, Phys. Rev. Lett. 89 (2002) 056402.
\newblock \href {http://dx.doi.org/10.1103/PhysRevLett.89.056402}
  {\path{doi:10.1103/PhysRevLett.89.056402}}.

\bibitem{gegenwart2008}
P.~Gegenwart, Q.~Si, F.~Steglich, Quantum cricritical in heavy fermion metals,
  Nature Physics 4 (2008) 186--197.
\newblock \href {http://dx.doi.org/10.1038/nphys892}
  {\path{doi:10.1038/nphys892}}.

\bibitem{custers2003}
J.~Custers, P.~Gegenwart, H.~Wilhelm, K.~Neumaier, Y.~Tokiwa, O.~Trovarelli,
  C.~Geibel, F.~Steglich, C.~P\'epin, P.~Coleman, The break-up of heavy
  electrons at a quantum critical point, Nature 424 (2003) 524--527.
\newblock \href {http://dx.doi.org/10.1038/nature01774}
  {\path{doi:10.1038/nature01774}}.

\bibitem{stock2012}
C.~Stock, C.~Broholm, F.~Demmel, J.~Van~Duijn, J.~Taylor, H.~Kang, R.~Hu,
  C.~Petrovic, From incommensurate correlations to mesoscopic spin resonance in
  $\mathrm{Y}\mathrm{b}\mathrm{R}{\mathrm{h}}_{\mathrm{2}}\mathrm{S}{\mathrm{i}}_{\mathrm{2}}$,
  Phys. Rev. Lett. 109 (2012) 127201.
\newblock \href {http://dx.doi.org/10.1103/PhysRevLett.109.127201}
  {\path{doi:10.1103/PhysRevLett.109.127201}}.

\bibitem{sichelschmidt2003}
J.~Sichelschmidt, V.~A. Ivanshin, J.~Ferstl, C.~Geibel, F.~Steglich, Low
  temperature electron spin resonance of the {K}ondo ion in a heavy fermion
  metal:
  $\mathrm{Y}\mathrm{b}\mathrm{R}{\mathrm{h}}_{\mathrm{2}}\mathrm{S}{\mathrm{i}}_{\mathrm{2}}$,
  Phys. Rev. Lett. 91 (2003) 156401.
\newblock \href {http://dx.doi.org/10.1103/PhysRevLett.91.156401}
  {\path{doi:10.1103/PhysRevLett.91.156401}}.

\bibitem{sichelschmidt2007}
J.~Sichelschmidt, J.~Wykhoff, H.-A. Krug~von Nidda, J.~Ferstl, C.~Geibel,
  F.~Steglich, Spin dynamics of spin dynamics of
  $\mathrm{Y}\mathrm{b}\mathrm{R}{\mathrm{h}}_{\mathrm{2}}\mathrm{S}{\mathrm{i}}_{\mathrm{2}}$
  observed by electron spin resonance, Journal of Physics: Condensed Matter 19
  (2007) 116204.
\newblock \href {http://dx.doi.org/10.1088/0953-8984/19/11/116204}
  {\path{doi:10.1088/0953-8984/19/11/116204}}.

\bibitem{wykhoff2007}
J.~Wykhoff, J.~Sichelschmidt, G.~Lapertot, G.~Knebel, J.~Flouquet, I.~I.
  Fazlishanov, H.-A. Krug~von Nidda, C.~Krellner, C.~Geibel, F.~Steglich, On
  the local and itinerant properties of the {ESR} in
  $\mathrm{Y}\mathrm{b}\mathrm{R}{\mathrm{h}}_{\mathrm{2}}\mathrm{S}{\mathrm{i}}_{\mathrm{2}}$,
  Science and Technology of Advanced Materials 8 (2007) 389--392.
\newblock \href {http://dx.doi.org/10.1016/j.stam.2007.07.005}
  {\path{doi:10.1016/j.stam.2007.07.005}}.

\bibitem{Duque2009}
J.~G.~S. Duque, E.~M. Bittar, C.~Adriano, C.~Giles, L.~M. Holanda,
  R.~Lora-Serrano, P.~G. Pagliuso, C.~Rettori, C.~A. P\'erez, R.~Hu,
  C.~Petrovic, S.~Maquilon, Z.~Fisk, D.~L. Huber, S.~B. Oseroff, Magnetic field
  dependence and bottlenecklike behavior of the {ESR} spectra in
  {YbRh}$_2${Si}$_2$, Phys. Rev. B 79 (2009) 035122.
\newblock \href {http://dx.doi.org/10.1103/PhysRevB.79.035122}
  {\path{doi:10.1103/PhysRevB.79.035122}}.

\bibitem{schaufuss2009}
U.~Schaufu\ss{}, V.~Kataev, A.~A. Zvyagin, B.~B\"uchner, J.~Sichelschmidt,
  J.~Wykhoff, C.~Krellner, C.~Geibel, F.~Steglich, Evolution of the {K}ondo
  state of
  $\mathrm{Y}\mathrm{b}\mathrm{R}{\mathrm{h}}_{\mathrm{2}}\mathrm{S}{\mathrm{i}}_{\mathrm{2}}$
  probed by high-field {ESR}, Phys. Rev. Lett. 102 (2009) 076405.
\newblock \href {http://dx.doi.org/10.1103/PhysRevLett.102.076405}
  {\path{doi:10.1103/PhysRevLett.102.076405}}.

\bibitem{Kochelaev2009}
B.~I. Kochelaev, S.~I. Belov, A.~M. Skvortsova, A.~S. Kutuzov,
  J.~Sichelschmidt, J.~Wykhoff, C.~Geibel, F.~Steglich, Why could electron spin
  resonance be observed in a heavy fermion {K}ondo lattice?, The European
  Physical Journal B 72 (2009) 485--489.
\newblock \href {http://dx.doi.org/10.1140/epjb/e2009-00386-9}
  {\path{doi:10.1140/epjb/e2009-00386-9}}.

\bibitem{Woelfle2009}
P.~W\"olfle, E.~Abrahams, Phenomenology of {ESR} in heavy-fermion systems:
  Fermi-liquid and non-fermi-liquid regimes, Phys. Rev. B 80 (2009) 235112.
\newblock \href {http://dx.doi.org/10.1103/PhysRevB.80.235112}
  {\path{doi:10.1103/PhysRevB.80.235112}}.

\bibitem{scheffler2013}
M.~Scheffler, K.~Schlegel, C.~Clauss, D.~Hafner, C.~Fella, M.~Dressel,
  M.~Jourdan, J.~Sichelschmidt, C.~Krellner, C.~Geibel, F.~Steglich, Microwave
  spectroscopy on heavy-fermion systems: Probing the dynamics of charges and
  magnetic moments, Physica Status Solidi {B} 250 (2013) 439--449.
\newblock \href {http://dx.doi.org/10.1002/pssb.201200925}
  {\path{doi:10.1002/pssb.201200925}}.

\bibitem{Javaheri2016}
M.~Javaheri~Rahim, T.~Lehleiter, D.~Bothner, C.~Krellner, D.~Koelle,
  R.~Kleiner, M.~Dressel, M.~Scheffler, Metallic coplanar resonators optimized
  for low-temperature measurements, J. Phys. D: Appl. Phys. 49 (2016) 1--6.
\newblock \href {http://dx.doi.org/10.1088/0022-3727/49/39/395501}
  {\path{doi:10.1088/0022-3727/49/39/395501}}.

\bibitem{Ghirri2015}
A.~Ghirri, C.~Bonizzoni, M.~Righi, F.~Fedele, G.~Timco, R.~Winpenny,
  M.~Affronte, Microstrip resonators and broadband lines for x-band epr
  spectroscopy of molecular nanomagnets, Applied Magnetic Resonance 46 (2015)
  749--756.
\newblock \href {http://dx.doi.org/10.1007/s00723-015-0672-5}
  {\path{doi:10.1007/s00723-015-0672-5}}.

\bibitem{Frunzio2005}
L.~Frunzio, A.~Wallraff, D.~Schuster, J.~Majer, R.~Schoelkopf, Fabrication and
  characterization of superconducting circuit qed devices for quantum
  computation, IEEE Transactions on Applied Superconductivity 15 (2005)
  860--863.
\newblock \href {http://dx.doi.org/10.1109/TASC.2005.850084}
  {\path{doi:10.1109/TASC.2005.850084}}.

\bibitem{Goeppl2008}
M.~G\"oppl, A.~Fragner, M.~Baur, R.~Bianchetti, S.~Filipp, J.~M. Fink, P.~J.
  Leek, G.~Puebla, L.~Steffen, A.~Wallraff, Coplanar waveguide resonators for
  circuit quantum electrodynamics, Journal of Applied Physics 104 (2008)
  113904.
\newblock \href {http://dx.doi.org/10.1063/1.3010859}
  {\path{doi:10.1063/1.3010859}}.

\bibitem{Clauss2013}
C.~Clauss, D.~Bothner, D.~Koelle, R.~Kleiner, L.~Bogani, M.~Scheffler,
  M.~Dressel, Broadband electron spin resonance from 500 {MHz} to 40 {GHz}
  using superconducting coplanar waveguides, Applied Physics Letters 102 (2013)
  162601.
\newblock \href {http://dx.doi.org/10.1063/1.4802956}
  {\path{doi:10.1063/1.4802956}}.

\bibitem{Malissa2013}
H.~Malissa, D.~I. Schuster, A.~M. Tyryshkin, A.~A. Houck, S.~A. Lyon,
  Superconducting coplanar waveguide resonators for low temperature pulsed
  electron spin resonance spectroscopy, Review of Scientific Instruments 84
  (2013) 025116.
\newblock \href {http://dx.doi.org/10.1063/1.4792205}
  {\path{doi:10.1063/1.4792205}}.

\bibitem{DiIorio1988}
M.~S. DiIorio, A.~C. Anderson, B.~Y. Tsaur, rf surface resistance of
  {Y-Ba-Cu-O} thin films, Phys. Rev. B 38 (1988) 7019--7022.
\newblock \href {http://dx.doi.org/10.1103/PhysRevB.38.7019}
  {\path{doi:10.1103/PhysRevB.38.7019}}.

\bibitem{Hafner2014}
D.~Hafner, M.~Dressel, M.~Scheffler, Surface-resistance measurements using
  superconducting stripline resonators, Rev. Sci. Instrum. 85 (2014) 014702.
\newblock \href {http://dx.doi.org/10.1063/1.4856475}
  {\path{doi:10.1063/1.4856475}}.

\bibitem{Thiemann2017}
M.~{Thiemann}, M.~H. {Beutel}, M.~{Dressel}, N.~R. {Lee-Hone}, D.~M. {Broun},
  E.~{Fillis-Tsirakis}, H.~{Boschker}, J.~{Mannhart}, M.~{Scheffler}, {Single
  gap superconductivity in doped {SrTiO}$_3$}, ArXiv e-prints\href
  {http://arxiv.org/abs/1703.04716} {\path{arXiv:1703.04716}}.

\bibitem{wiemann2015}
Y.~Wiemann, J.~Simmendinger, C.~Clauss, L.~Bogani, D.~Bothner, D.~Koelle,
  R.~Kleiner, M.~Dressel, M.~Scheffler, Observing electron spin resonance
  between 0.1 and 67 {GHz} at temperatures between 50 {mK} and 300 {K} using
  broadband metallic coplanar waveguides, Appl. Phys. Lett. 106 (2015) 193505.
\newblock \href {http://dx.doi.org/10.1063/1.4921231}
  {\path{doi:10.1063/1.4921231}}.

\bibitem{Scheffler2015}
M.~Scheffler, M.~M. Felger, M.~Thiemann, D.~Hafner, K.~Schlegel, M.~Dressel,
  K.~S. Ilin, M.~Siegel, S.~Seiro, C.~Geibel, F.~Steglich, Broadband corbino
  spectroscopy and stripline resonators to study the microwave properties of
  superconductors, Acta IMEKO 4 (2015) 47.
\newblock \href {http://dx.doi.org/10.21014/acta_imeko.v4i3.247}
  {\path{doi:10.21014/acta_imeko.v4i3.247}}.

\bibitem{Parkkinen2015}
K.~Parkkinen, M.~Dressel, K.~Kliemt, C.~Krellner, C.~Geibel, F.~Steglich,
  M.~Scheffler, Signatures of phase transitions in the microwave response of
  {YbRh}$_2${Si}$_2$, Physics Procedia 75 (2015) 340--347.
\newblock \href {http://dx.doi.org/10.1016/j.phpro.2015.12.040}
  {\path{doi:10.1016/j.phpro.2015.12.040}}.

\bibitem{Voesch2015}
W.~Voesch, M.~Thiemann, D.~Bothner, M.~Dressel, M.~Scheffler, On-chip {ESR}
  measurements of {DPPH} at m{K} temperatures, Physics Procedia 75 (2015)
  503--510.
\newblock \href {http://dx.doi.org/10.1016/j.phpro.2015.12.063}
  {\path{doi:10.1016/j.phpro.2015.12.063}}.

\bibitem{Attocube}
Attocube ANRv51/RES.

\bibitem{krellner2012}
C.~Krellner, S.~Taube, T.~Westerkamp, Z.~Hossain, C.~Geibel, Single-crystal
  growth of {YbRh}$_2${Si}$_2$ and {YbIr}$_2${Si}$_2$, Philosophical Magazine
  92 (2012) 2508--2523.
\newblock \href {http://dx.doi.org/10.1080/14786435.2012.669066}
  {\path{doi:10.1080/14786435.2012.669066}}.

\bibitem{Bothner2012a}
D.~Bothner, T.~Gaber, M.~Kemmler, D.~Koelle, R.~Kleiner, S.~W\"unsch,
  M.~Siegel, Magnetic hysteresis effects in superconducting coplanar microwave
  resonators, Phys. Rev. B 86 (2012) 014517.
\newblock \href {http://dx.doi.org/10.1103/PhysRevB.86.014517}
  {\path{doi:10.1103/PhysRevB.86.014517}}.

\bibitem{deGraaf2012}
S.~E. de~Graaf, A.~V. Danilov, A.~Adamyan, T.~Bauch, S.~E. Kubatkin, Magnetic
  field resilient superconducting fractal resonators for coupling to free
  spins, Journal of Applied Physics 112 (2012) 123905.
\newblock \href {http://dx.doi.org/10.1063/1.4769208}
  {\path{doi:10.1063/1.4769208}}.

\bibitem{Bothner2012b}
D.~Bothner, C.~Clauss, E.~Koroknay, M.~Kemmler, T.~Gaber, M.~Jetter,
  M.~Scheffler, P.~Michler, M.~Dressel, D.~Koelle, R.~Kleiner, Reducing vortex
  losses in superconducting microwave resonators with microsphere patterned
  antidot arrays, Applied Physics Letters 100 (2012) 012601.
\newblock \href {http://dx.doi.org/10.1063/1.3673869}
  {\path{doi:10.1063/1.3673869}}.

\bibitem{Ebensperger2016}
N.~G. Ebensperger, M.~Thiemann, M.~Dressel, M.~Scheffler, Superconducting {Pb}
  stripline resonators in parallel magnetic field and their application for
  microwave spectroscopy, Supercond. Sci. Technol. 29 (2016) 115004.
\newblock \href {http://dx.doi.org/10.1088/0953-2048/29/11/115004}
  {\path{doi:10.1088/0953-2048/29/11/115004}}.

\bibitem{Degiorgi1999}
L.~Degiorgi, The electrodynamic response of heavy-electron compounds, Rev. Mod.
  Phys. 71 (1999) 687--734.
\newblock \href {http://dx.doi.org/10.1103/RevModPhys.71.687}
  {\path{doi:10.1103/RevModPhys.71.687}}.

\bibitem{Scheffler2005c}
M.~Scheffler, M.~Dressel, M.~Jourdan, H.~Adrian, Extremely slow {D}rude
  relaxation of correlated electrons, Nature 438 (2005) 1135.
\newblock \href {http://dx.doi.org/10.1038/nature04232}
  {\path{doi:10.1038/nature04232}}.

\bibitem{Scheffler2010}
M.~Scheffler, M.~Dressel, M.~Jourdan, Microwave conductivity of heavy fermions
  in {UPd}$_2${Al}$_3$, The European Physical Journal B 74 (2010) 331--338.
\newblock \href {http://dx.doi.org/10.1140/epjb/e2010-00085-6}
  {\path{doi:10.1140/epjb/e2010-00085-6}}.

\bibitem{joshi2004}
J.~P. Joshi, S.~Bhat, On the analysis of broad {D}ysonian electron paramagnetic
  resonance spectra, Journal of Magnetic Resonance 168 (2004) 284--287.
\newblock \href {http://dx.doi.org/10.1016/j.jmr.2004.03.018}
  {\path{doi:10.1016/j.jmr.2004.03.018}}.

\end{thebibliography}

\end{document}